\documentclass[11pt]{article}
\usepackage[utf8]{inputenc}
\usepackage{geometry}
\usepackage{amssymb}
\usepackage{upgreek}
\usepackage{setspace}
\usepackage{amsmath}
\usepackage{makecell}
\usepackage{multirow}
\usepackage{adjustbox}
\usepackage{url}
\usepackage{algorithm}
\usepackage{algpseudocode}
\newcommand{\mb}{\mathbf}
\newcommand{\bs}{\boldsymbol}

\usepackage{verbatim}
\usepackage{caption}

\usepackage[compact]{titlesec}
\titlespacing{\section}{0pt}{2ex}{1ex}
\titlespacing{\subsection}{0pt}{1ex}{0ex}
\titlespacing{\subsubsection}{0pt}{0.5ex}{0ex}

\usepackage{natbib}
\usepackage{graphicx}

\usepackage{xr}
\makeatletter
\newcommand*{\addFileDependency}[1]{
  \typeout{(#1)}
  \@addtofilelist{#1}
  \IfFileExists{#1}{}{\typeout{No file #1.}}
}
\makeatother

\begin{document}

\thispagestyle{empty} \baselineskip=28pt \vskip 5mm
\begin{center}
{\Huge{\bf  A Physics-Informed, Deep Double Reservoir Network for Forecasting Boundary Layer Velocity}}
\end{center}

\baselineskip=12pt \vskip 10mm

\begin{center}\large
Matthew Bonas\footnote[1]{
\baselineskip=11pt Department of Applied and Computational Mathematics and Statistics,
University of Notre Dame, Notre Dame, IN 46556, USA.}, David H. Richter\footnote[2]{Department of Civil and Environmental Engineering and Earth Sciences,
University of Notre Dame,  Notre Dame, IN 46556, USA.} and Stefano Castruccio\textsuperscript{1}
\end{center}

\baselineskip=17pt \vskip 10mm \vskip 10mm

\begin{center}
{\large{\bf Abstract}}
\end{center}

\doublespacing

When a fluid flows over a solid surface, it creates a thin boundary layer where the flow velocity is influenced by the surface through viscosity, and can transition from laminar to turbulent at sufficiently high speeds. Understanding and forecasting the fluid dynamics under these conditions is one of the most challenging scientific problems in fluid dynamics. It is therefore of high interest to formulate models able to capture the nonlinear spatio-temporal velocity structure as well as produce forecasts in a computationally efficient manner. Traditional statistical approaches are limited in their ability to produce timely forecasts of complex, nonlinear spatio-temporal structures which are at the same time able to incorporate the underlying flow physics. In this work, we propose a model to accurately forecast boundary layer velocities with a deep double reservoir computing network which is capable of capturing the complex, nonlinear dynamics of the boundary layer while at the same time incorporating physical constraints via a penalty obtained by a Partial Differential Equation (PDE). Simulation studies on a one-dimensional viscous fluid demonstrate how the proposed model is able to produce accurate forecasts while simultaneously accounting for energy loss. The application focuses on boundary layer data in a water tunnel with a PDE penalty derived from an appropriate simplification of the Navier-Stokes equations, showing improved forecasting by the proposed approach in terms of mass conservation and variability of velocity fluctuation against non-physics-informed methods.

\newpage

\section{Introduction}\label{sec:Intro}

\quad Understanding the dynamics and properties of the fluid boundary layer --- the thin region of flow near a solid surface where the flow is influenced by the boundary --- requires the incorporation of many processes that occur across a range of temporal (i.e., seconds to hours) and spatial (i.e., centimeters to hundreds of meters) scales \citep{stu88}. These boundary layer dynamics are well-studied in the field of fluid mechanics but are complex and highly relevant to a range of applications in both engineering and atmospheric science, including wind turbine energy production \citep{cli14}, air pollution transport and dispersion \citep{fer10}, agriculture \citep{fer10b} and urban habitability \citep{con15}. 

In the context of this work, ``fluid'' refers broadly to any medium (i.e., liquid, gas, or plasma) that is governed by a continuum description of motion and response to forcing. As such, the velocities of interest behave as a nonlinear spatio-temporal process whose dynamics can be described by the Navier-Stokes equations, which encode fundamental physical constraints based on conservation of mass and momentum. Recent progresses in machine learning have prompted a different, data-driven approach to modeling and forecasting (see, e.g., \cite{huang95, lei09, ak16, kho18, demo19, huang2021}). Such methods are predicated on the use of highly non-parametric constructs such as deep Neural Networks (NN) for modeling the space-time dynamics, with increasingly accurate results. Despite these successful implementations, this class of models brings new challenges and questions related to the forecasts.
\begin{enumerate}
    \item \textit{Practicality}. Since inference on deep NN is notoriously time consuming, its operational use is hampered by computational challenges. Is is possible to produce flexible yet computationally affordable deep NNs for fast boundary layer forecasting?
    \item \textit{Interpretability}. The use of purely data driven methods discards any physical information already available for the problem, in this case the Navier-Stokes equations. Is it possible to leverage on the considerable flexibility of deep NN, while producing forecasts which are at least approximately compliant with the basic laws of fluid dynamics? 
\end{enumerate}

As far as practicality is concerned, numerous recent approaches have been developed to allow computationally faster NN in time (\textit{recurrent NNs}). At the core of these \textit{reservoir computing methods} is the acknowledgement that weight matrices in NN with unknown entries are excessively parametrized, and that these matrices could be instead regarded as sparse realizations from a stochastic model. Among these methods, \textit{echo-state networks} (ESNs, \cite{jae01, jae07}) are arguably the most popular of them. Recent works \citep{mcd17, mcd19a, mcd19b} showed the ability of this model to capture the dynamics of nonlinear spatio-temporal processes while also properly assessing the uncertainty in the forecasts. Other more recent works have applied ESNs to predict complex environmental processes such as air pollution transmission \citep{bonas23}, wind energy forecasting \citep{huang2021}, fire front propagation \citep{yoo23}, and smoke plume identification \citep{lar21}. Reservoir computing methods have also been combined with the spiking neuron approach to create stochastic methods able to model binary sequences or spike streams with a deep NN \citep{maass97}. These \textit{liquid state machines} (LSMs, \cite{maass02}) have been used in prior works in the areas of forecasting sea clutter data \citep{ross16}, wind speed \citep{wei21} and exchange-traded funds \citep{maten21}.

For the second challenge (interpretability), a recent branch of machine learning has focused on constraining parameter learning in NNs to be more compliant with physical constraints, expressed as Partial Differential Equations (PDEs). These \textit{physics-informed neural networks} (PINNs, \cite{rais19, zhang19, meng20, pang20, north22}) assume the existence of additional constraints arising from symmetries, invariance, or conservation laws from the context of the scientific investigation, which are then used as additional constraints in the function to minimize, just as any penalized inference approach such as ridge regression and lasso \citep{has09}. Despite the considerable computational overhead, modern computers have enabled the computation of its gradient and hence perform gradient-based optimization (backpropagation) in an affordable manner \citep{bay18}. PINNs bear resemblance with functional spatial regression with PDE penalization, a popular area in statistics predicated on the same principles in a functional setting with a continuous spatial domain \citep{san12,san21}.

In this work, we aim at tackling the challenges of practicality and interpretability of forecasting boundary layer velocity by means of a new model that leverages all of the aforementioned features and merges them in a single framework. Specifically, we define a hybrid DESN and DLSM model (deep double reservoir network) which is able to leverage both continuous and binary input. We rely on a physics-informed penalized minimization with a PDE penalty, and we present a computationally affordable algorithm to achieve inference. Finally, we propose a calibration step to achieve proper uncertainty quantification in the forecasts. In the simulation study, we use a simplified one-dimensional viscous fluid, and we show how the proposed approach produces improved forecasts against both statistical and machine learning methods for space-time data. Additionally, our method is able to predict a consistent energy loss in the system, a feature that a data-driven models cannot achieve. Finally, in the application we consider high resolution laboratory observations of a turbulent boundary layer, and penalize its forecast with a simplified form of the Navier-Stokes equations. The resulting model produces accurate forecasts which are also physically consistent and well calibrated. From a practical point of view, this work is meant to serve as a foundation for producing physically-informed and realistic forecasts of atmospheric data and simulations.

The manuscript proceeds as follows. In Section \ref{sec:Data} we introduce the boundary layer water data which will be used in the application. Section \ref{sec:methods} describes the DESN, the DLSM, the double reservoir model, and the physics-informed penalization. Section \ref{sec:simstudy} presents a simulation study on a one dimensional viscous fluid that assesses the performance of our approach against alternative statistical and machine learning models, in terms of point prediction as well as physical consistency of the predictions. Section \ref{sec:Appl} applies the model to the boundary layer velocity and shows how the forecasts are compliant with mass conservation. Section \ref{sec:conclusion} concludes with a discussion.

\section{Data}\label{sec:Data}

\quad We consider experimentally obtained velocity data of a turbulent boundary layer from time-resolved particle image velocimetry (PIV, \cite{data15, fisc18}), across a spatial field for over 2,500 time steps at frequency of 1,000 Hz, so that the collection window is 2.5s. The data are collected from an experimental water tunnel where the underlying space-time process can be described by the incompressible Navier-Stokes equations, which will be used as physics-based penalization in Section \ref{sec:Infer}. This type of laboratory experiment --- where the working fluid is water instead of air --- is a prototypical example of the concept of \textit{dynamic similarity} in fluid mechanics. According to this principle, it is the relative strengths of the dominant physical mechanisms which matters, and not the individual, dimensional fluid properties. A common example of this is the dimensionless Reynolds number, which compares the strength of fluid inertia to the viscous forces. Flows of air or water with the same Reynolds number are considered dynamically similar, enabling the use of scale models for studying flows at different dimensional scales (e.g., wind tunnel studies of large-scale flows). In the current context, the water tunnel experiments represent a scale model of the atmospheric boundary layer, where motions on the order of seconds in the tunnel represent motions on the order of minutes or hours in the atmospheric boundary layer.

These data were originally obtained to provide high temporal and spatial resolution of the boundary layer, to better understand the spectral energy characteristics of the near-wall turbulence \citep{data15, fisc18}. Under the principle of dynamic similarity, in which flows are found to be self-similar at some length and time scales, these data also serve as a scale model of the types of motions one would observe in, for example, the atmosphere over minutes or hours. The data are collected on a Cartesian grid of approximate size $0.19$m$\times$$0.05$m, where the points are equally spaced with horizontal ($x$) and vertical ($y$) resolution of 0.00037m; overall, there are over 62,000 spatial locations with two-dimensional velocity data. The bulk flow characteristics are provided as part of the experiment and include the boundary layer thickness, $\delta = 0.1m$; friction velocity, $u_{\tau} = 0.027m s^{-1}$; and Reynolds number, $Re_{\tau} = 2,700$. This Reynolds number is large, allowing the tunnel experiments to physically represent a wide range of flows (including the atmospheric boundary layer) where fluid inertia strongly dominates over the viscous forces trying to dissipate kinetic energy.

\begin{figure}[!tb]
\centering
\includegraphics[width = 12cm]{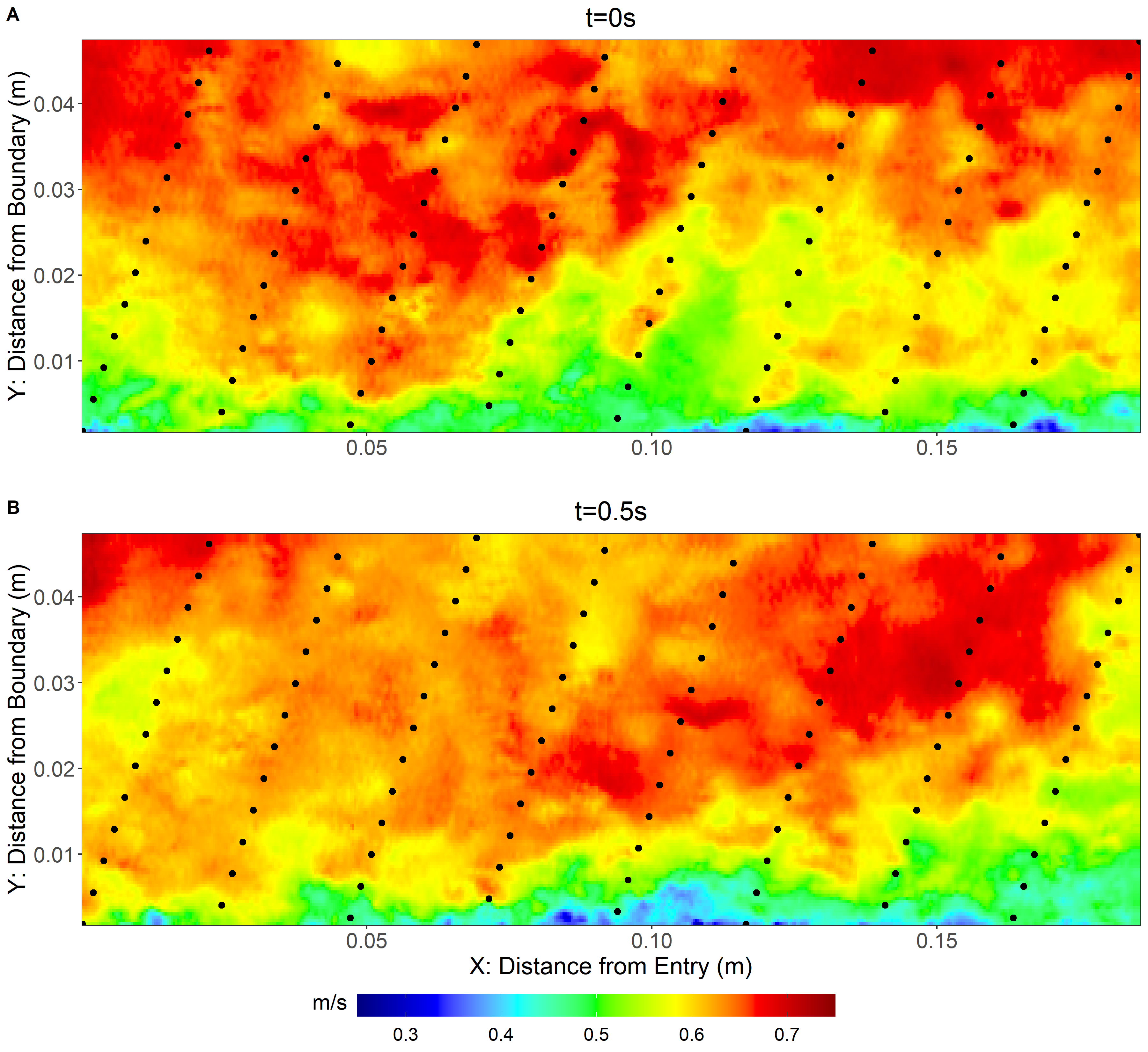}
\caption{Water speed in m/s for the water field data from \cite{data15, fisc18} at the initial time (panel A) and at $t=0.5$s (panel B). The black dots represent the 100 subsampled locations used for analysis and forecasting.}
\label{fig:WF_Data}
\end{figure}

In this controlled water field experiment we use each of these parameters when implementing the physics-informed penalization using a time-averaged version of the Navier-Stokes equations in Section \ref{sec:Appl}. The data were filtered to a frequency of 200 Hz (aggregated in time by a factor of 5$\times$), so that the total number of time points is now $T=500$. Additionally, we subsampled $N=100$ points on a equally spaced diagonal grid from the original spatial locations as shown in Figure \ref{fig:WF_Data}. The choice of a diagonal grid allows us to retain the important information in the vertical ($y$) direction since the vertical spacing between locations is small ($<$0.004m or 8\% of the entire vertical spacing) while also allowing us to offset some possible minor patterns existing in the $x$ direction. The forecast for the $N$ locations is interpolated with thin plate splines \citep{green94} to obtain the remaining spatial domain. Figure \ref{fig:WF_Data} shows the entire spatial domain for both the initial time point (panel A) and at $t=0.5s$ (panel B). The black points in these panels represent the $N$ subsampled locations, which will be used for point prediction, physical constraints, and uncertainty quantification in Section \ref{sec:Appl}. Throughout this work, we denote the velocity field by $\mb{Y}_{t}=(Y_t(\mb{s}_1),\ldots, Y_t(\mb{s}_N))^\top$ at locations $\mb{s}_1, \ldots, \mb{s}_N$ and times $t=1, \ldots, T$. In the simulation study we employ $\mb{Y}_{t}$ directly, while in the application we focused on a time averaged version.

\section{Methodology}\label{sec:methods}

\quad We introduce a DESN in Section \ref{sec:ESN} and the DLSM in Section \ref{sec:SNN}. In Section \ref{sec:PINNSCombo} we propose the physics-informed, double-reservoir model used throughout this work. We also present the physics-informed inference in Section \ref{sec:Infer}. The forecast calibration and uncertainty quantification are deferred to the supplementary material.

\subsection{Deep Echo-State Networks}\label{sec:ESN}

\quad ESNs \citep{jae01, jae07} are stochastic modifications of a recurrent NN assuming that the weights connecting the input to the hidden states and the weights interconnecting the hidden states to themselves are drawn from some prior distribution controlled by some parameters. The introduction of stochastic weight matrix avoids any inherent instabilities associated with gradient based optimization algorithms (i.e., backpropagation) of recurrent NNs. Additionally, instead of estimating any weight as a parameter, the ESN implies that a small parametric space can be optimized using a cross-validation approach, thereby reducing the computational cost. For this work, we use a deep ESN (DESN, \cite{mcd19b}), a model predicated on sampling the weights within the network multiple times and averaging the corresponding forecasts. The DESN will be used as a reference model for comparison throughout the remainder of this work.

A DESN is formulated as follows \citep{mcd19b}:

\begin{subequations}\label{eqn:DESN}
\begin{flalign}
\text{output:} & \qquad \mb{Y}_{t} = \mb{B}_{D}\mb{h}_{t, D} + \sum_{d=1}^{D-1} \mb{B}_{d}k\left(\Tilde{\mb{h}}_{t,d}\right) + \bs{\epsilon}_{t} \label{eqn:DESN1}, \\
\text{hidden state $d$:} & \qquad \mb{h}_{t,d} = (1-\alpha)\mb{h}_{t-1,d} + \alpha\bs{\omega}_{t,d} \label{eqn:DESN2},\\
& \qquad \bs{\omega}_{t,d} = f_{h}\left(\frac{\zeta_{d}}{|\lambda_{W_{d}}|} \mb{W}_{d} \mb{h}_{t-1,d} + \mb{W}_{d}^{\text{in}}\Tilde{\mb{h}}_{t,d-1}\right), \text{ for } d > 1 \label{eqn:DESN3},\\
\text{reduction $d-1$:} & \qquad \Tilde{\mb{h}}_{t,d-1} \equiv Q\left(\mb{h}_{t,d-1}\right), \text{ for } d > 1 \label{eqn:DESN4},\\
\text{input:} & \qquad \bs{\omega}_{t,1} = f_{h}\left(\frac{\zeta_{1}}{|\lambda_{W_{1}}|} \mb{W}_{1} \mb{h}_{t-1, 1} + \mb{W}_{1}^{\text{in}}\mb{x}_{t}\right) \label{eqn:DESN5},\\
\text{matrix distribution:} & \qquad W_{d_{i,j}} = \gamma^{W_{d}}_{i,j} p(\eta_{W_{d}}) + (1-\gamma^{W_{d}}_{i,j})\delta_{0} \label{eqn:DESN6},\\
& \qquad W^{\text{in}}_{d_{i,j}} = \gamma^{W_{d}^{\text{in}}}_{i,j} p(\eta_{W_{d}^{\text{in}}}) + (1-\gamma^{W_{d}^{\text{in}}}_{i,j})\delta_{0} \label{eqn:DESN7},\\
& \qquad \gamma^{W_{d}}_{i,j} \text{\;$\thicksim$\;} Bern(\pi_{W_{d}}), \quad \gamma^{W_{d}^{\text{in}}}_{i,j} \text{\;$\thicksim$\;} Bern(\pi_{W_{d}^{\text{in}}}). \nonumber
\end{flalign}
\end{subequations}

\noindent Here, $d = \{1,\dots,D\}$ represents the depth or the number of layers in the network. The output $\mb{Y}_{t}$ is expressed as a linear combination $\mb{B}_{D}\mb{h}_{t, D} + \sum_{d=1}^{D-1} \mb{B}_{d}k_{h}\left(\Tilde{\mb{h}}_{t,d}\right) + \bs{\epsilon}_{t}$ in \eqref{eqn:DESN1}, while $\mb{h}_{t,D}$ is an $n_{h,D}$-dimensional state vector, $\Tilde{\mb{h}}_{t,d}$ is an $n_{\Tilde{h},d}$-dimensional state vector, and $\mb{B}_{d}$ are matrices with unknown entries. The error term $\bs{\epsilon}_{t}$ in \eqref{eqn:DESN1} is assumed to be independent identically distributed in time and is denoted as a multivariate normal distribution with zero mean. The state vector is represented as a convex combination of its past state, $\mb{h}_{t-1,d}$, and a memory term, $\bs{\omega}_{t,d}$, controlled by the hyper-parameter $\alpha$ (known as the `leaking-rate') in \eqref{eqn:DESN2}. The dimension of the hidden state $\mb{h}_{t,d}$ is reduced using the function $Q(\cdot)$ in equation \eqref{eqn:DESN4}. This results in the $n_{\Tilde{h}, d}$-dimensional hidden state $\Tilde{\mb{h}}_{t,d}$, and in this work, we use an empirical orthogonal function \citep{hann07} approach as a dimension reduction function $Q(\cdot)$. The $k(\cdot)$ in equation \eqref{eqn:DESN1} represents some scaling function needed to ensure that the values of $\Tilde{\mb{h}}_{t,d}$ are on a similar scale to those within $\mb{h}_{t,D}$ (i.e., the range of values which the two vectors take is similar or the same) and in this work we chose the hyperbolic tangent function. In equations \eqref{eqn:DESN3} and \eqref{eqn:DESN5}, the term $\bs{\omega}_{t,d}$ is derived from a nonlinear function (\textsl{activation function}) $f_{h}$ with some pre-specified forms such as the hyperbolic tangent or rectified linear unit \citep{goo16}. This function combines the past hidden state $\mb{h}_{t-1,d}$ and some layer specific input data. This input data is specified as either the $n_{x}$-dimensional vector of input $\mb{x}_{t}$, which comprises of the past/lagged values of $\mb{Y}_{t}$, (i.e.,  $\mb{x}_{t}=\left(\mb{Y}_{t-\tau},\ldots,\mb{Y}_{t-m\tau}\right)^\top$, where $\tau$ represents the forecast lead time \citep{mcd17}) and $m \ge 1$ represents the total number of lags of the lead-time to retain, or the dimension reduced hidden state from the prior layer $\Tilde{\mb{h}}_{t,d-1}$. A spike-and-slab prior is used to inform the entries of the weight matrices $\mb{W}_{d}$ and $\mb{W}^{\text{in}}_{d}$ in \eqref{eqn:DESN6} and \eqref{eqn:DESN7} where individual entries of these matrices take a value of zero with some given probability $\pi_{W_{d}}$ and $\pi_{W_{d}^{\text{in}}}$ for $\mb{W}_{d}$ and $\mb{W}^{\text{in}}_{d}$, respectively. The non-zero elements of the weight matrices are drawn from some symmetric distribution centered about zero \citep{luk12} and for this work we assume $p(\cdot) \sim Unif(-1,1)$, however other choices of distribution have been shown to be effective choices in previous works (see \cite{mcd17, huang2021, bonas23, yoo23}). Lastly, the DESN must also respect the \textit{echo-state property}, i.e. after a sufficiently long time sequence, the model must asymptotically lose its dependence on the initial conditions \citep{luk12,jae07}. This property holds when the spectral radius (largest eigenvalue) of $\mb{W}_{d}$, denoted by $\lambda_{W_{d}}$, is less than one. The scaling parameter $\zeta_{d} \in (0,1]$ is introduced to ensure that the spectral radius does not violate this condition on equation \eqref{eqn:DESN3} and \eqref{eqn:DESN5}. Following \cite{mcd19b}, we fix the hyper-parameters $\pi_{W_{d}}$ and $\pi_{W_{d}^{\text{in}}}$ to 0.1, as the results have been shown to be robust with respect to this choice. We assume that $n_{h,d}=n_h$ and $n_{\Tilde{h}, d}=n_{\Tilde{h}}$, and we denote the model hyper-parameters by $\bs{\theta}_{ESN}=\{n_{h}, n_{\Tilde{h}}, m, \{\zeta_{d}, d=1, \ldots, D\},\lambda_{r}, \alpha, D\}$, where $\lambda_{r}$ is the penalty for ridge regression as will be described in Section \ref{sec:Infer}. From the parameter list we thereby exclude $\mb{B}_d, d=1, \ldots, D$, which are estimated separately as will be discussed in Section \ref{sec:Infer}.

\subsection{Deep Liquid State Machines}\label{sec:SNN}

\quad Traditional machine learning models such as feedforward NNs compute and transmit continuous valued signals throughout the network. Unlike these, spiking NNs rely on discrete spike streams (binary sequences) to process information, and were formulated to more closely mimic natural biological NNs \citep{maass97}. In this work, we formulate a dynamic version of a spiking NN, a DLSM \citep{maass02}, which discretizes the input through a leaky integrate-and-fire model (LIF, \citet{burk06, zhang18}), which we detail in Section \ref{sec:LIFs}. In Section \ref{sec:LSMs} we provide the mathematical description of the DLSM. The notation in Section \ref{sec:LIFs} will be used exclusively for introducing the LIF model, whereas in Section \ref{sec:LSMs} a slightly different notation will be used, in order to align it with the DESN in the previous Section. Additional details about the heuristic behind this model can be found in the supplementary material. 

\subsubsection{Preliminaries: The Leaky Integrate-and-Fire Model}\label{sec:LIFs}

\quad The LIF models the dynamics of $\mb{s}_{t} = \{{s}^{(i)}_{t}; i = 1, \dots, n_{h}\}$, a $n_{h}$-dimensional binary discrete time series whose input is another $(m+1)$-dimensional binary discrete time series $\mb{I}_t=\{I^{(j)}_{t}; j = 1, \dots, m+1\}$. Heuristically, the said input impacts the \textit{membrane potential energy} $\bs{V}_{\text{mem}}(t) = \{{V}^{(i)}_{\text{mem}}(t); i = 1, \dots, n_{h}\}$, and once it crosses a threshold $V_{\text{thr}}$ (which in this work is the same for each neuron $i$) it allows the output to be one, i.e., $s^{(i)}_{t} = 1$. Once the threshold is passed, the potential is set to a resting value $V_{\text{res}}$, which we fix to zero for all $i$. Finally, during each time step, a small portion of the membrane potential energy is `leaked' out, and the overall potential membrane energy is decreased by a value $V_{\text{leak}}$, again identical for each $i$. 

Formally, for each neuron $i$ the LIF model is defined as follows:
\begin{subequations}\label{eqn:SNNPrelims}
\begin{gather}
{V_{\text{mem}}^{(i)}(t)} = \sum_{j \in \Phi}w_{j;i}I_t^{(j)} - V_{\text{leak}} + V^{(i)}_{\text{mem}}(t-1), \quad \text{if} \quad V^{(i)}_{\text{mem}}(t) < V_{\text{thr}},\label{eqn:SNNPrelims1}\\
V^{(i)}_{\text{mem}}(t) = V_{\text{res}}, \quad \text{if} \quad V^{(i)}_{\text{mem}}(t) \ge V_{\text{thr}} \label{eqn:SNNPrelims2},\\
s^{(i)}_{t} = \left\{\begin{array}{lr}
                    1, & \text{if } V^{(i)}_{\text{mem}}(t) \ge V_{\text{thr}},\\
                    0, & \text{if } V^{(i)}_{\text{mem}}(t) < V_{\text{thr}},
                   \end{array}
            \right. \label{eqn:SNNPrelims3}
\end{gather}
\end{subequations}

\noindent where $\Phi$ represents the set of synapses for which the input impacts the membrane potential, $w_{j;i}$ are the synapse weights for neuron $i$ associated with the input spike stream $I_t^{(j)}$ which is either 1 or 0. As will be discussed in Section \ref{sec:LSMs}, DLSMs adhere to the same principle as the DESN involving the random and sparse generation of the synapse weights and $\Phi$ here represents the set of synapses which connect the input to the neurons which have non-zero weight. The membrane potential is updated using equation \eqref{eqn:SNNPrelims1}, compared against the threshold in equation \eqref{eqn:SNNPrelims2} and possibly reset to the resting state. The individual elements of the output are defined in equation \eqref{eqn:SNNPrelims3}.

\subsubsection{Mathematical Definition of Liquid State Machines}\label{sec:LSMs}

\quad A DLSM is a reservoir computing model based on the same principle of random sparse matrices as the DESN in equations \eqref{eqn:DESN3} and \eqref{eqn:DESN5}, adapted to the case of spiking NNs. Here, for each time point $t$ we assume a dynamic model on a sub-time scale $t^*=1, \ldots, T^{*}$, and we count the number of output spikes generated by a LIF model. More formally, we redefine $\bs{\omega}_{t, d}$ from equations \eqref{eqn:DESN3} and \eqref{eqn:DESN5} as the total number of spikes from the LIF model in Section \ref{sec:LIFs}:
\[
\bs{\omega}_{t, d} = \sum_{t^{*}=1}^{T^{*}}\mb{s}_{t^*, d},
\]

\noindent where $\bs{\omega}_{t, d}$ is a vector of size $n_{h}$ as in the DESN model in equation \eqref{eqn:DESN}. We further specify the evolution of the vector of the neurons' potential energy $\bs{V}_{\text{mem}}(t^{*})$ during each sub-time point, $t^{*}$ by reformulating equations \eqref{eqn:DESN3} and \eqref{eqn:DESN5} as follows:
\begin{subequations}\label{eqn:MemUpdate}
\begin{gather}
\Delta \bs{V}_{\text{mem}}(t^{*}) = \frac{\zeta}{|\lambda_{W}|} \mb{W} \mb{s}_{t^{*}-1} + \mb{W}^{\text{in}}\Tilde{\mb{x}}_{t^{*}} - V_{\text{leak}} -  \text{max}(\mb{s}_{t^{*}}) V_{\text{inhib}}, \\
\Delta \bs{V}_{\text{mem}}(t^{*}) = 0 \quad \text{if} \quad  \mb{s}_{t^{*}-\Tilde{t}} = 1, \quad \forall \Tilde{t} \in [0, V_{\text{latent}}], 
\end{gather}
\end{subequations}

\noindent Here, $\Tilde{\mb{x}}_{t^{*}}$ represents the input spike stream, $\bs{I}_{t^*}=(I^{(j)}_{t^*}; j = 1, \ldots, n_{x})$, from equation \eqref{eqn:SNNPrelims}. We allow the neurons to go through a latency period $V_{\text{latent}}$ after emitting a spike to emulate the natural delayed response from the stimulus of the spike \citep{lek15}. This latency period forces the neurons' membrane potential energy to neither increase nor decay over a fixed number of sub-time points. We also allow each neuron to exhibit lateral inhibition \citep{Cohen11}, or the ability of neurons to reduce the membrane potential energy of the other neurons in the same layer when they emit a spike. Lateral inhibition is implemented by reducing the membrane potential energy of every neuron in the hidden state by some fixed amount, $V_{\text{inhib}}$, when one neuron emits a spike. The total number of hyper-parameters for the DLSM are then $\bs{\theta} = \{\bs{\theta}_{ESN}, V_{\text{thr}}, V_{\text{leak}}, V_{\text{inhib}}, V_{\text{latent}}\}$, where $\bs{\theta}_{ESN}$ is the same hyper-parameter vector as in the DESN.

In order to feed data into the DLSM, the continuous input we use in our application must be converted into spike trains. To convert the input data, $\mb{x}_{t}$ for the first layer and $\Tilde{\mb{h}}_{t,d}$ for all subsequent layers, from equations \eqref{eqn:DESN3} and \eqref{eqn:DESN5} into spike trains, the data must first be rescaled using min-max normalization, formally:
\begin{eqnarray}\label{eqn:MinMaxNormal}
\mb{x}_{t}^{\prime} = \frac{\mb{x}_{t} - \text{min}(\mb{x}_{t})}{\text{max}(\mb{x}_{t}) - \text{min}(\mb{x}_{t})} \nonumber.
\end{eqnarray}

\noindent The $n_{x}$-length vector $\mb{x}_{t}^{\prime}$ is then converted to a matrix of spike trains, $\Tilde{\mb{x}}_{t^{*}}$, of dimension $n_{x} \times T^{*}$, by creating spike trains of length $T^{*}$ with firing rates proportional to the scaled vector elements of $\mb{x}_{t}^{\prime}$. This approach resembles spike train conversion in image classification, where the gray levels and color intensities of a pixel are converted into spike trains with firing rates proportional to their pixel value \citep{wije19}, hence implying that the values are drawn from a Poisson process \citep{oconnor13, cao14, diehl15}. 
The use of a sub-time scale $T^{*}$ implies a considerable increase in the length of the observational vector. For example in our work we choose $T^{*}=100$, and the conversion into spike trains converts a network with $n=1,000$ observations to one with $n\times T^{*}= 100,000$ observations. The specification of $T^{*}$ is solely dictated by the computational resources available, as a small value would allow faster computation but lower accuracy of the spike train in representing the signal, while a larger value than the one proposed would make the computation practically impossible. 


\subsection{Deep Double-Reservoir Network}\label{sec:PINNSCombo}

\quad In this work, we combine a DESN and a DLSM in a deep double reservoir network (DDRN) to improve the flexibility and the predictability of the underlying process. Formally, the DDRN reformulates the output of the model in equation \eqref{eqn:DESN1} as:
\begin{subequations}\label{eqn:ComboOutput}
\begin{flalign}
& \qquad \mb{Y}_{t} =  \mb{Z}^{\text{(DESN)}}_{t} + \mb{Z}^{\text{(DLSM)}}_{t} + \bs{\epsilon}_{t}, \\
& \qquad \mb{Z}^{\text{(DESN)}}_{t} = \mb{B}^{\text{(DESN)}}_{D}\mb{h}^{\text{(DESN)}}_{t, D} + \sum_{d=1}^{D-1} \mb{B}^{\text{(DESN)}}_{d}k\left(\Tilde{\mb{h}}^{\text{(DESN)}}_{t,d}\right), \\
& \qquad \mb{Z}^{\text{(DLSM)}}_{t} = \mb{B}^{\text{(DLSM)}}_{D}\mb{h}^{\text{(DLSM)}}_{t, D} + \sum_{d=1}^{D-1} \mb{B}^{\text{(DLSM)}}_{d}k\left(\Tilde{\mb{h}}^{\text{(DLSM)}}_{t,d}\right), 
\end{flalign}
\end{subequations}

\noindent which assumes a linear combination of the hidden states (reservoirs) from the DESN and DLSM to explain the output $\mb{Y}_{t}$. The model coefficients $\mb{B}=\{\mb{B}_{d}^{\text{(DESN)}}, \mb{B}_{d}^{\text{(DLSM)}}, d=1, \ldots, D\}$ are unknown and need to be estimated as detailed in the next Section.

\subsection{Inference with Physics-Based Networks}\label{sec:Infer}

\quad A portion of the observed training data is held out for validation and used to perform inference for the hyper-parameter vector $\bs{\theta}$. For a fixed $\bs{\theta}$ the set of parameters comprising of the collection of all matrices $\mb{B}$ from equation \eqref{eqn:ComboOutput} is estimated according to the method described in this Section. The predictions are then produced and the resulting mean squared error (MSE) is minimized with respect to the set of hyper-parameters in $\bs{\theta}$. This approach has been shown to generate successful hyper-parameter estimates for the DESN (see \cite{mcd17, bonas23, bonas23b} for some examples) and relies on the data being stationary in time, an assumption which is true by design in our application on boundary layer velocity. In other applications where stationarity is not realistic, one could use on an out-of-sample validation approach across multiple testing periods for parameter optimization \citep{cerq20}.

The simplest approach to estimate the coefficients in $\mb{B}$ is via a matrix version of ridge regression with some penalty $\lambda_{r}$ (see the supplementary material for a detailed description), which is computationally affordable and prevents exceedingly large estimator variance by adding some bias \citep{mcd17,huang2021,bonas23, yoo23}. In this work, we operate under the assumption that additional context is available about the underlying spatio-temporal process in the form of a PDE, and it is of interest to integrate it in the analysis by penalizing the inference towards it. Indeed, beyond producing accurate point forecasts with our proposed model, we are also interested in ensuring that the forecast at least loosely follows the physical laws characterizing the underlying spatio-temporal process. To ensure this, instead of using an $L^2$ penalty as in the case of ridge regression, the loss function is modified to include a penalty quantifying the extent of the departure of the estimated data to the physical equation. Formally, if we denote by $Y_t(\mb{s})$ the time varying continuous field of interest, we assume we are given a spatio-temporal PDE in the form of \citep{rais19}:
\begin{eqnarray}\label{eqn:nonlinPDE}
\frac{\partial {Y}_t(\mb{s})}{\partial t} + N[Y_t(\mb{s})]\nonumber=0,
\end{eqnarray}

\noindent where $N[Y_t(\mb{s})]$ is a (possibly nonlinear) differential operator, which in this work is assumed to be known. For example, in the case of the viscous Burgers' equation in Section \ref{sec:burger}, the process is one dimensional in space so that $\mb{s}=x$ and the operator is defined as $N[Y_t(x)] = \xi_{1}Y_t(x)\frac{\partial Y_t(x)}{\partial x} - \xi_{2}\frac{\partial^2 Y_t(x)}{\partial x^2}$, with $\xi_{1}$ and $\xi_{2}$ assumed to be known. 

Instead of enforcing this equation strictly, we only penalize the departure of the solution from the equation. Formally, for the estimated $\hat{Y}_t(\mb{s})$, we define  
\[
\hat{g}_t(\mb{s})=\frac{\partial \hat{Y}_t(\mb{s})}{\partial t} + N[\hat{Y}_t(\mb{s})]
\]
so that in the case of a perfect physics-obeying solution we have $\hat{g}_t(\mb{s}) = 0$ for all $\mb{s}$ and $t$. We then construct a new penalty which trades off the prediction skills with the magnitude of the aforementioned quantify \citep{rais19}:
\begin{eqnarray}\label{eqn:ModifiedMSE}
&& \text{MSE}_{\text{pred}+\text{phys}} = (1-\chi)\text{MSE}_{\text{pred}} + \chi \text{MSE}_{\text{phys}}, \\
&& \text{MSE}_{\text{pred}} = \frac{1}{NT}\sum_{i=1}^{N}\sum_{t=1}^{T}|{Y}_t(\mb{s}_{i}) - \hat{{Y}}_{t}(\mb{s}_{i})|^2,\nonumber \\
&& \text{MSE}_{\text{phys}} = \frac{1}{NT}\sum_{i=1}^{N}\sum_{t=1}^{T}|\hat{{g}}_t(\mb{s}_i)|^2,\nonumber 
\end{eqnarray}

\noindent where $\text{MSE}_{\text{pred}}$ is the MSE for the point forecasts to the observed data $Y_t(\mb{s})$ and $\text{MSE}_{\text{phys}}$ corresponds to the error respective to the PDE. The parameter $\chi \in [0,1]$ represents the relative weight of the predictive MSE with respect to the physical constraints, which we consider as an additional hyper-parameter. Unlike the ridge regression case, there is no closed-form for either the solution nor the gradient of $\text{MSE}_{\text{pred}+\text{phys}}$. As such, given the very large number of parameters with respect to which the optimization is sought. The sheer number of parameters needed to be optimized with this newly defined loss function ultimately creates a significant computational bottleneck when it comes to optimization. To mitigate the computational cost, we first optimize the PIDESN model using the initial estimates from ridge regression (assuming the loss function is defined as MSE$_{\text{pred}}$, which has a closed form expression). We then initialize the PIDDRN optimization by setting the parameters associated with the DESN component to these estimates, while the parameters of the DLSM component are set to zero. This approach then guarantees that the PIDDRN performs at least as well as the PIDESN. Despite this solution, inference for all entries of $\mb{B}$ is still computationally infeasible, so to further reduce the computational cost, we optimize only a proportion $\psi \in [0,1]$ of the matrix $\mb{B}$ weights (a sensitivity study with respect to the choice of parameters is performed in Table S1), while the remaining $\lfloor (1-\psi)|\mb{B}|\rfloor$ weights are fixed. In Figure S1 we show how inference for larger values of $\psi$ would be computationally infeasible.

\section{Simulation Study}\label{sec:simstudy}

\quad We aim to assess the predictive capabilities of the PIDDRN model against the non physics-informed DDRN, the DESN, and physics-informed DESN (PIDESN), as well as other time series and machine learning approaches. In this section, we focus on simulated data from the viscous Burgers' equation \citep{burger48}, which we introduce in Section \ref{sec:burger}. We show a comparison in terms of their predictive ability and uncertainty quantification in Section \ref{sec:burgerForcsUQ}.

Reservoir computing methods such as DESN have been shown to produce computationally affordable forecasts owing to the use of random sparse matrices \citep{mcd17, huang2021, bonas23, yoo23}, and the PIDDRN is no exception. Even with the addition of the DLSM and the physics-informed loss function, inference can still be achieved in a reasonable time. For this simulation study, we were able to produce 500 ensemble forecasts using the PIDDRN in parallel using only 24 cores at 2.50Ghz frequency in less than 12 hours. This approach could easily be up-scaled by increasing the number of cores. The computational time could also be decreased by using GPUs instead of CPUs.

\subsection{Viscous Burgers' Equation}\label{sec:burger}

\begin{figure}[!tb]
\centering
\includegraphics[width = 14cm]{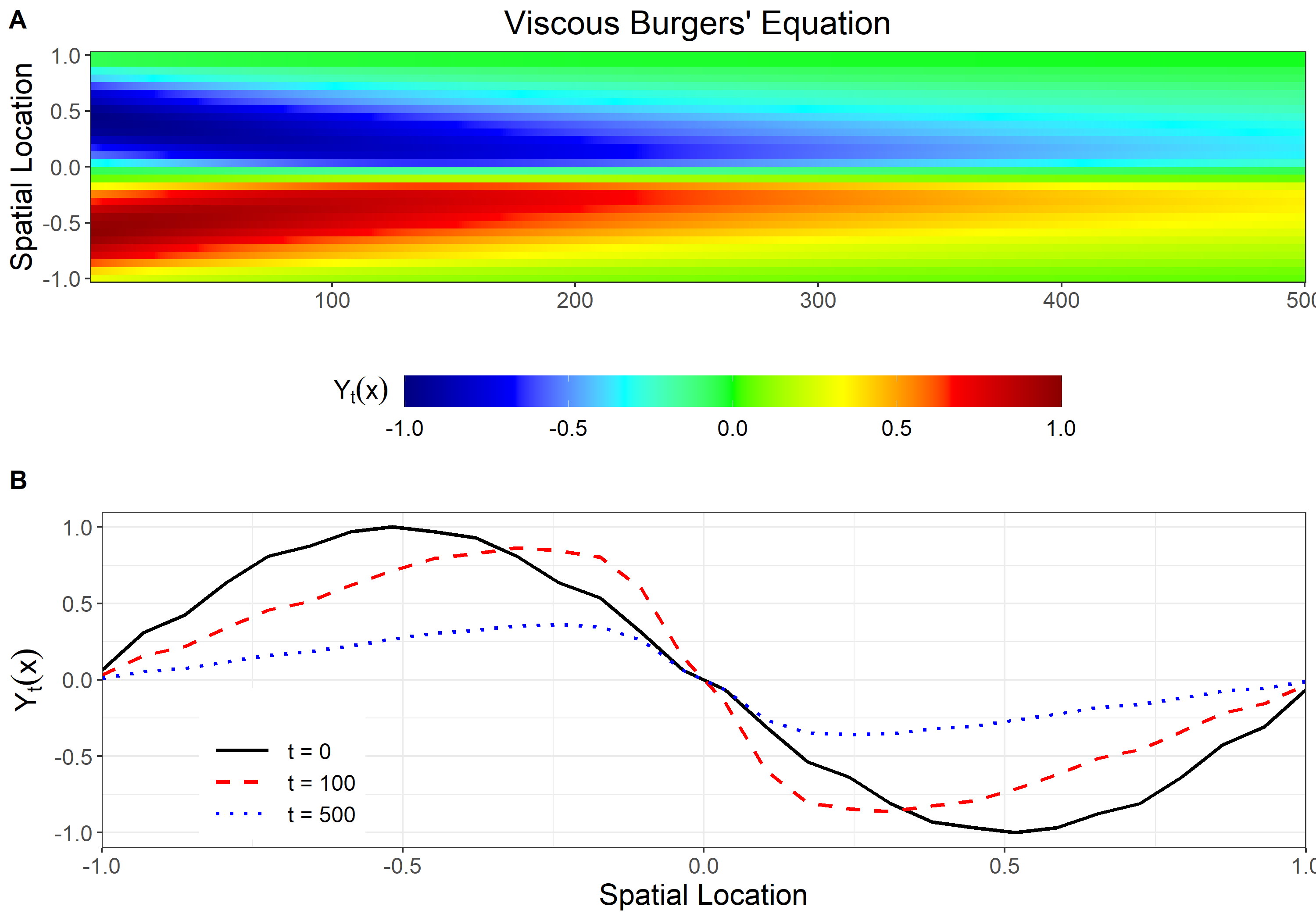}
\caption{Solution of the viscous Burgers' equation from equation \eqref{eqn:Burgers} with initial condition $Y_0(x) = -\sin(\pi x)$ and viscosity $\nu = 0.01$. (A) Heatmap across all points in time ($x$ axis) and space ($y$ axis). (B) Solution for spatial points, for three time points $t=0$, $t=100$ and $t=500$.}
\label{fig:BurgerSpaceTime}
\end{figure}

\quad The viscous Burgers' equation is often employed as a simple surrogate for the Navier-Stokes equations. Here we take the one-dimensional form which describes the space-time velocity of a fluid of a given constant viscosity $\nu>0$ \citep{bate15, burger48}. Formally, it is defined as:
\begin{equation}\label{eqn:Burgers}
\frac{\partial Y_t(x)}{\partial t}  +  Y_t(x)\frac{\partial Y_t(x)}{\partial x} = \nu\frac{\partial^2 Y_t(x)}{\partial x^2}
\end{equation}

\noindent where $Y_t(x)$ is the fluid velocity at some location $x$ and time point $t$. The Burgers' equation operates in continuous time, but as assume we observe data at equally spaced times using a discretized version of it. Indeed, we simulate data at $N=30$ uniformly distributed spatial locations on $x \in [-1,1]$ across 500 equally spaced time steps. We solve the equation using a spatial discretization of 1/15 and temporal discretization of 3/1000 (so a total time of 1.5), a choice which allows for a numerically stable solution to the PDE. We specify two different initial conditions: a sinusoid $Y_0(x) = -\sin(\pi x)$ and a Gaussian curve $Y_0(x) = e^{-x^{2}}$. Additionally, we consider viscosity values of $\nu = \left\{10^{-s}; s = 1, \dots, 4\right\}$, for a total of 8 different simulation setups. The simulated data with $Y_0(x) = -\sin(\pi x)$ and $\nu = 0.01$ can be seen in Figure \ref{fig:BurgerSpaceTime}A, as well as the velocity $Y_t(x)$ for all $x$ at times $t=0$, $t=100$, and $t=500$ in Figure \ref{fig:BurgerSpaceTime}B. Each model was trained for each of the simulations using the first $T=450$ time steps and was tested using the next $\tau=50$ steps. We produced forecasts using 1000 ensembles, each representing an independent draw for the weight matrices $\mb{W}_{d}$ and $\mb{W}^{\text{in}}_{d}$ in equations \eqref{eqn:DESN6},  \eqref{eqn:DESN7} and \eqref{eqn:MemUpdate}. The grids used with the validation approach to optimize the hyper-parameter space are shown in the supplementary material in Table S4. In the supplementary material we further show the ability of the proposed PIDDRN model to produce accurate long lead forecasts on a generalized Burgers' equation with an external forcing \citep{buy13}.

\subsection{Point Forecasts and Uncertainty Quantification}\label{sec:burgerForcsUQ}

\quad We compare the forecasting performance of PIDDRN not only with the DDRN and DESN models described above, but also with a physics-informed version of the DESN (PIDESN) and other time series and machine learning approaches. We choose an auto-regressive, fractionally integrated moving average model (ARFIMA, \cite{granger80,hosking81}) as well as a standard machine learning approach for time series, the long short-term memory model (LSTM, \cite{hoch97}). This model is a special case of a recurrent NN, but relies on fully connected, dense weight matrices which are optimized via gradient-based methods (technical details are provided in the supplementary material). Finally, we compare the results with a na\"ive temporal forecasting approach (persistence) which assumes the $\tau$ future forecasts assume the value of the final observed data in the training set.

The forecasting skills are assessed in terms of the MSE$_{\text{pred}+\text{phys}}$ metric described in equation \eqref{eqn:ModifiedMSE} and the results are shown in Table \ref{tbl:BurgerMethodComp}, with the median performance across all viscosity values and the interquartile range (IQR) in parenthesis. It can easily be seen how the PIDDRN dramatically outperforms all other models. More specifically, the PIDDRN model returns a median MSE$_{\text{pred}+\text{phys}}$ for the sinusoidal initial condition of 0.04 (with IQR 0.002) and for the Gaussian initial condition of 0.08 (0.07). The results show an improvement against the DESN model of 89.7\% and 63.6\% for the sinusoidal and Gaussian initial conditions, respectively. The PIDDRN also shows a dramatically improved prediction against ARFIMA, whose median MSEs of 0.69 (0.99) and 0.36 (0.10) indicates significantly worse performance for both initial conditions. Additionally, the PIDDRN model outperforms its non-physics-informed counterpart, the DDRN, which has a smaller improvement over the standard DESN of returning MSE values of 0.33 (0.61) and 0.14 (0.08) for the two initial conditions, respectively. Finally, the PIDDRN model strongly outperforms the physics-informed version of the reference DESN, the PIDESN, which itself only obtained a modest improvement in MSE, returning values of 0.23 (0.36) and 0.19 (0.05).

\begin{table}[!tb]
\begin{centering}
\begin{tabular}{||c|c|c||} 
\hline
Forecasting Method & $Y_0(x) = -\text{sin}(\pi x)$ & $Y_0(x) = e^{-x^2}$ \\ [1ex] 
\hline\hline
PIDDRN & 0.04 (0.002) & 0.08 (0.07) \\
\hline
DDRN & 0.33 (0.61) & 0.14 (0.08) \\
\hline
PIDESN & 0.23 (0.36) & 0.19 (0.05) \\
\hline
DESN & 0.39 (0.70) & 0.22 (0.09) \\
\hline
ARFIMA & 0.69 (0.99) & 0.36 (0.10) \\
\hline
Long Short-Term Memory & 0.33 (0.57)  & 0.20 (0.05) \\
\hline
Persistence (Na\"ive) & 0.36 (0.70)  & 0.10 (0.06) \\
\hline
\end{tabular}
\caption{Forecasting performance for each model in terms of the MSE$_{\text{pred}+\text{phys}}$ metric described in equation \eqref{eqn:ModifiedMSE}. A median across all simulated datasets (choices of $\nu$) for each initial condition is shown with the IQR in parenthesis.}
\label{tbl:BurgerMethodComp}
\end{centering}
\end{table}

\begin{figure}[!tb]
\centering
\includegraphics[width = 14cm]{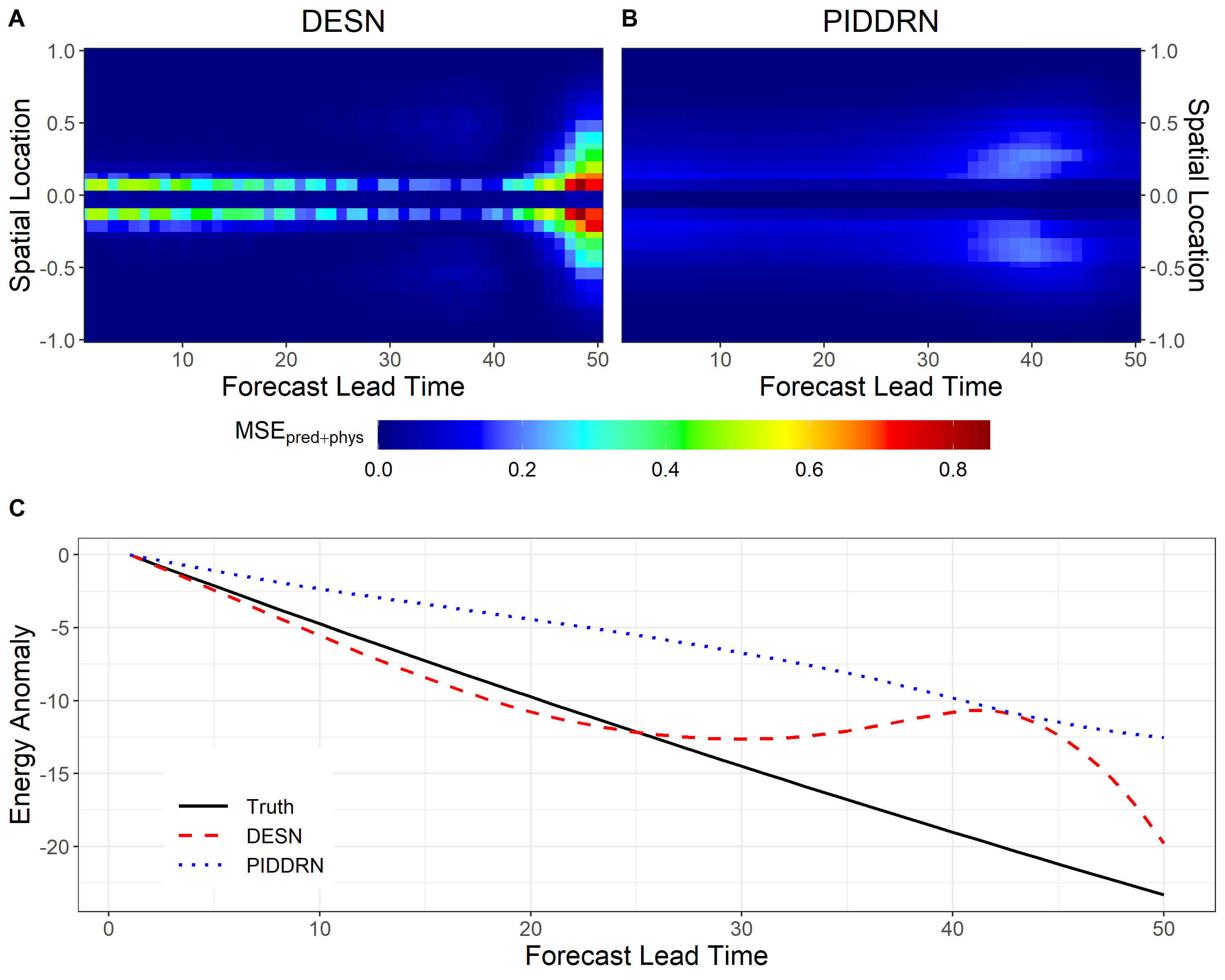}
\caption{Prediction comparison in terms of the MSE$_{\text{pred}+\text{phys}}$ from \eqref{eqn:ModifiedMSE} for the viscous Burgers' equation \eqref{eqn:Burgers}. The first row (panels A and B) represents data from the simulation where $Y_0(x) = -\text{sin}(\pi x)$ and $\nu = 0.01$. The second row panels (C) represents the total energy profile (in terms of energy anomaly with respect to the initial forecast time point) of the system across space through time, for both the DESN and PIDDRN models.}
\label{fig:Burger_MSE}
\end{figure}

To further illustrate the improvement of the PIDDRN model, the spatio-temporal maps of MSE$_{\text{pred}+\text{phys}}$ for the non-physics-informed DESN and physics-informed PIDDRN model for the initial condition $Y_0(x) = -\text{sin}(\pi x)$ with viscosity $\nu = 0.01$ can be seen in Figure \ref{fig:Burger_MSE}(A-B). It can be seen how the non-physics-informed DESN produce almost uniformly (in space and time) worse results than PIDDRN. The relative ordering of improvement among the models from Table \ref{tbl:BurgerMethodComp} is clearly demonstrated in this figure, where it is shown how the target prediction metric progressively improves, with PIDDRN being able to effectively capture the spatio-temporal areas which have forecasting inadequacies in DESN. 

The Burgers' equation assumes a one dimensional fluid excited by an initial condition, progressively decaying through the action of the viscosity and hence gradually losing energy. In order to assess the physical consistency of PIDDRN forecasts against DESN, in Figure \ref{fig:Burger_MSE}(C) we compare the total energy of the system, calculated as the squared integral in space for each time point, as a function of forecast lead time (as anomalies from the last observed time point). It can be seen how the PIDDRN model correctly predicts a monotonically decreasing energy, while DESN, being not informed about the physical properties of the equation, produces an nonphysical energy increase, especially at large lead times. 

The calibration method discussed in the supplementary material was implemented to produce uncertainty estimates surrounding the PIDDRN forecasts. We choose to compare the uncertainty estimates generated via this method (see supplementary material or \cite{bonas23b} for further details) versus the ensemble based approach introduced in \cite{mcd17}. This ensemble based approach is predicated on the normality assumption for the data, and the standard deviation of the forecasts ensemble is used to generate the uncertainty estimates. Table \ref{tbl:BurgerUQ} shows a comparison of the marginal uncertainty quantification when the forecasts are uncalibrated (i.e., using the estimates of the standard deviations from the ensemble of forecasts, \cite{mcd17}) and calibrated using the aforementioned approach. This table shows that, when using the uncalibrated estimates, the marginal uncertainty is incorrectly quantified, thereby resulting in a drastically smaller coverage of the nominal 95\%, 90\%, and 80\% prediction intervals (PIs). Indeed, the average discrepancy in coverage from the uncalibrated estimates across all simulations is 33.5\% across the three intervals. When using the uncertainty estimates produced by the calibration approach, the marginal uncertainty is more correctly quantified and does capture an appropriate amount of the true data within the specified intervals. In fact, the calibrated 95\% PI now captures a median of 92.9\% of the true data across both space and time for all simulations and only produce a coverage discrepancy of 1.6\% across the three intervals.

\begin{table}[t!]
\begin{centering}
\begin{tabular}{||c|c|c||} 
\hline
Prediction Interval & Uncalibrated & Calibrated \\[1ex] 
\hline\hline
95\% & 69.4 (13.0) & 92.9 (5.3) \\ 
\hline
90\% & 55.8 (11.0) & 92.1 (7.3) \\ 
\hline
80\% & 39.4 (10.5) & 80.7 (5.8) \\
\hline
\end{tabular}
\caption{Uncertainty quantification methods compared in terms of their nominal 95\% PI coverage on data from the Burgers' equation \eqref{eqn:Burgers}. The ensemble-based (uncalibrated) approach is outlined in \cite{mcd17} and the quantile regression (calibrated) approach is the method detailed in \cite{bonas23b}. The median coverage across all simulations for both initial conditions as well as the IQR (in parenthesis) is reported.}
\label{tbl:BurgerUQ}
\end{centering}
\end{table}

\section{Application}\label{sec:Appl}

\quad We now apply our proposed PIDDRN model to forecast the boundary layer velocity for the application presented in Section \ref{sec:Data}. The first $T=490$ time steps were used for training whereas the remaining $\tau = 10$ points were used as the testing data, and we produced forecasts for 1000 ensembles. The final 10 points in the training data were used as the validation set to optimize the hyper-parameters for the models, and the grid search is identical to that used for the Burgers' equation in Section \ref{sec:simstudy}. The hyper-parameters which were deemed optimal for the reservoir computing models are shown in the supplementary material.

For this application we assume $\chi = 0.5$ in equation \eqref{eqn:ModifiedMSE}, i.e., the predictive error and the physical constraints are equally weighted. The derivation of the PDE penalty from the incompressible Navier-Stokes equations is relatively involved so for the sake of  simplicity it is deferred to the supplementary material, we only report the final result with some heuristic explanation. The main simplifying assumption is that the time-averaged field is one-dimensional; i.e., the time-averaged velocity component in the $x$ direction is nonzero, while in the $y$ direction ($\bar{v}$) and $z$ directions ($\bar{w}$) are equal to zero everywhere. This also implies the derivatives with respect to the $x$ and $z$ direction are equal to zero. Also, since we assume the flow is at equilibrium (\textsl{fully developed}), we neglect the derivative in the $x$ direction as well, and in our application the velocity $Y_t(\mb{s}), \mb{s}=(x,y)$ reduces to a scalar component in the $x$ direction as a function of $x$ and $y$. The forecasts are spatio-temporal, but they are then averaged in time and denoted as $\bar{Y}=\bar{Y}(x,y)$, and the PDE penalty is expressed with respect to this quantity:

\begin{equation}\label{eqn:AppPDE}
    \frac{u_{\tau}^{2}}{\delta}~ + \frac{u_{\tau}\delta}{\text{Re}_{\tau}} \frac{\partial^{2} \bar{Y}}{\partial y^{2}} + \left\{(\kappa y)^{2}\left| \frac{\partial \bar{Y}}{\partial y} \right|\right\} \frac{\partial \bar{Y}}{\partial y}~ =0,
\end{equation}.

\noindent where $\kappa$ is defined in the supplementary material and $\delta, u_{\tau}, \text{Re}_{\tau}$ are parameters defined in Section \ref{sec:Data}. The PDE in \eqref{eqn:AppPDE} will be used to define the function $\hat{g}$ in \eqref{eqn:ModifiedMSE} (which here is only a spatial function) and the target function to minimize.

We show a comparison of each of the models in terms of predictive ability in Section \ref{sec:ApplForc} and in terms of uncertainty quantification in Section \ref{sec:ApplUQ}. 

\subsection{Point Forecasts}\label{sec:ApplForc}

\begin{figure}[!tb]
\centering
\includegraphics[width = 14cm]{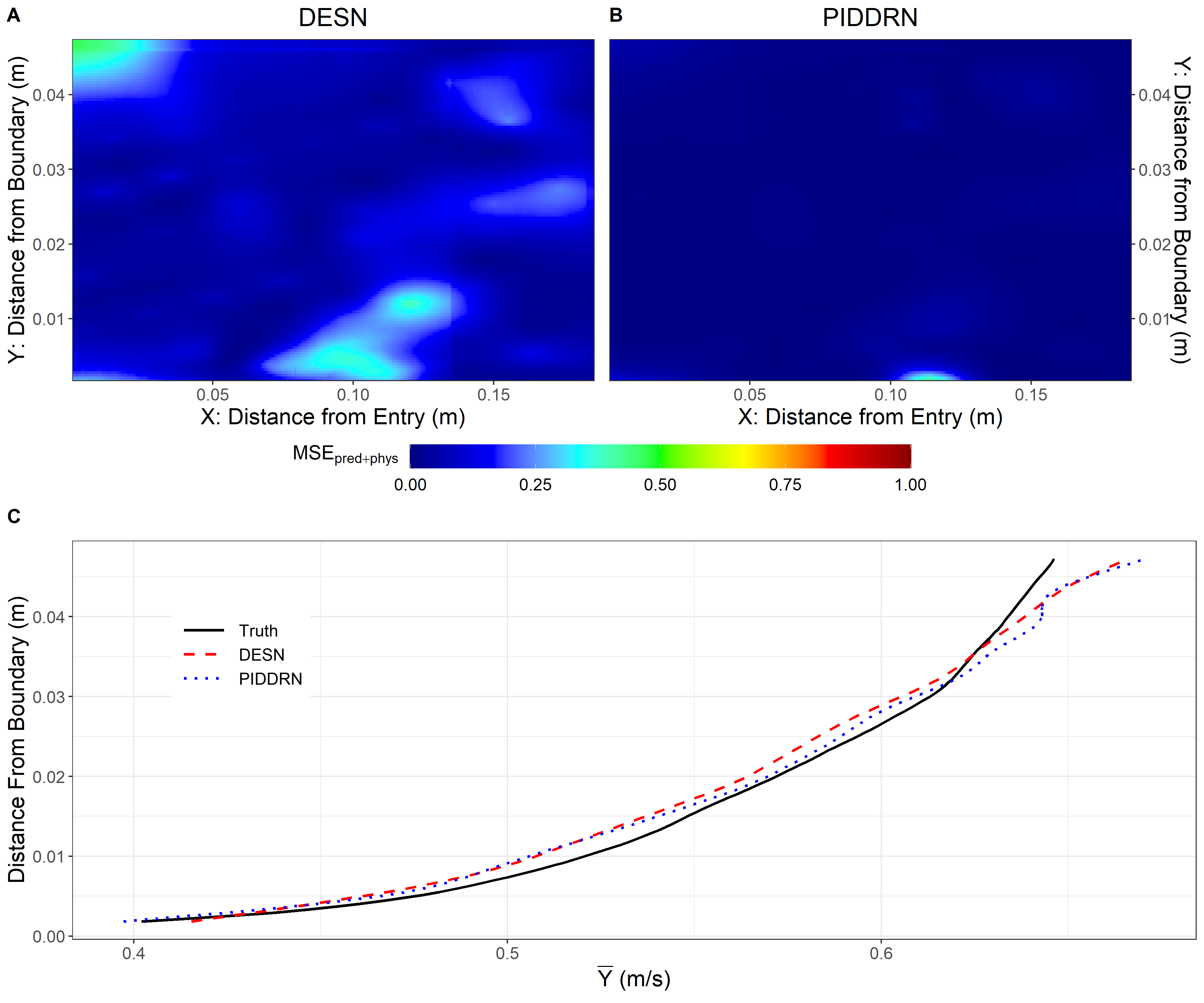}
\caption{A comparison of the MSE$_{\text{pred}+\text{phys}}$ from equation \eqref{eqn:ModifiedMSE} across the entire space time domain of the time-averaged water field data for the DESN (panel A) and PIDDRN (panel B) models. Panel C shows the estimated vertical profiles for the forecasts from both the DESN and PIDDRN models, averaged both in time and across $x$. `Distance from Entry' represents the distance in meters from the entrance or opening to the water tunnel used for the experiment and `Distance from Boundary' represents the distance in meters from the lower tunnel wall.}
\label{fig:WF_MSE}
\end{figure}

\quad We will show the forecasting performance of the PIDDRN model in terms of the physics-based loss for the 100 subsampled locations referenced in Section \ref{sec:Data} against the other similar reservoir computing models, the time series approach ARFIMA, the machine learning approach LSTM and the na\"ive forecasting approach persistence in terms of the time average of the forecast $\bar{Y}$. Similar to Table \ref{tbl:BurgerMethodComp}, Table \ref{tbl:WFMethodComp} shows the MSE$_{\text{pred}+\text{phys}}$ metric for each model for the forecast on the test data. As it is shown, this table illustrates how the PIDDRN again outperforms each of the other methods presented. In fact, the PIDDRN model produces a median MSE of 0.133 (with IQR of 0.046) which is an improvement over the DESN of 50.2\% which produces values for the MSE of 0.267 (0.168), a result further illustrated in Figure \ref{fig:WF_MSE}. Additionally, the PIDDRN model has a more significant improvement over the next best models, the DDRN and PIDESN models, where they only produce a MSE of of 0.262 (0.167) and 0.241 (0.130), respectively. As in the case of the simulation study, the DDRN does show improvement against the DESN model, yet it still under-performs relative its physics-informed counterpart, the PIDDRN. Finally, the time series model ARFIMA and machine learning model LSTM are the only models in this instance to produce worse results in terms of the MSE versus the DESN approach, for they produce MSEs of 0.296 (0.121) and 0.275 (0.127), respectively.

The spatio-temporal maps of the forecasting MSE$_{\text{pred}+\text{phys}}$ for the entire domain for the DESN and PIDDRN models can be seen in Figure \ref{fig:WF_MSE}(A-B). The forecasts from the 100 locations were used to interpolate the entire spatio-temporal domain using thin plate splines \citep{green94} and then these interpolated forecasts are used to calculate the MSE$_{\text{pred}+\text{phys}}$ across the entire space, as shown in Figure \ref{fig:WF_MSE}. Since we rely on a time-averaged simplification of the Navier-Stokes equations, the said loss is also time-independent, thus the panels represent this time-averaged loss for the two models. From these panels it can be seen how the PIDDRN produces remarkably better forecasting performance versus the DESN based models, a result consistent with those presented in Table \ref{tbl:WFMethodComp}. The DESN model fails to produce entirely accurate forecasts for the spatial domain in terms of the point prediction and physical constraints. 

We also compare the PIDDRN forecasts against the DESN and the data through the estimated vertical profile in panel C of Figure \ref{fig:WF_MSE}. In this case, both models estimate the true time- and $x$-averaged profile well, however a case could be made that the PIDDRN predictions are more physically consistent. Indeed, in our application and according to our derivation in the supplementary material the velocity divergence reduces to $\frac{\partial \bar{Y}}{\partial x} = 0$, so that mass conservation can be estimated by the extent to which this term is close to 0. Under this metric the PIDDRN model outperforms the other models, for this term returns an average value of 1.14 (with spatial standard deviation of 24.0) whereas the DESN models return 1.72 (21.2). This metric for comparison is ultimately a byproduct of the PDE penalization and the model itself is not being constrained to optimize this metric directly. However, since the proposed model is constrained to produce more physically consistent forecasts, this result lends some further support to the proposed PIDDRN model versus the DESN and DDRN.

To further show the physical consistency of the PIDDRN forecasts versus the DESN, we can also compute the variance of the fluctuation of the results around their time average, for each of the forecasts produced from their respective models. Given that we are penalizing the time-averaged velocity $\bar{u}$, we expect these variances to more closely follow those produced from the data, despite the penalization not being directly applied to the instantaneous velocity $u_{t}$. Thus we can simply calculate the MSE between the true fluctuation variances and the forecast fluctuations. In this instance, we observe that the DESN model produces an average MSE in time (with standard deviation in parenthesis) of 0.01 (0.003) whereas the PIDDRN model returns a much lower value of 0.002 (0.001). As before, the PDE penalty does not directly ensure smaller variation, this result is just indirect evidence of a more physical forecast, and it is of high interest in engineering applications as understanding turbulent fluctuations in time allows to assess fatigue in mechanical parts (e.g., wind turbine blades).

\begin{table}[!tb]
\begin{centering}
\begin{tabular}{||c|c||} 
\hline
Forecasting Method & MSE$_{\text{pred}+\text{phys}}$\\ [1ex] 
\hline\hline
PIDDRN & 0.133 (0.046) \\
\hline
DDRN & 0.262 (0.167) \\
\hline
PIDESN & 0.241 (0.130) \\
\hline
DESN & 0.267 (0.168) \\
\hline
ARFIMA & 0.296 (0.121) \\
\hline
Long Short-Term Memory & 0.275 (0.127) \\
\hline
Persistence (Na\"ive) & 0.267 (0.119) \\
\hline
\end{tabular}
\caption{Forecasting performance for each model in terms of the median MSE$_{\text{pred}+\text{phys}}$ (IQR in parenthesis) across the 100 subsampled locations described in Section \ref{sec:Data} for the water field application data.}
\label{tbl:WFMethodComp}
\end{centering}
\end{table}

\subsection{Uncertainty Quantification}\label{sec:ApplUQ}

\quad As was the case with the simulated data in Section \ref{sec:simstudy}, the calibration method discussed in the supplementary material was implemented with the PIDDRN model to produce uncertainty estimates surrounding the forecasts. In this instance we chose a sliding window of $20$ time points as part of the implementation. Table \ref{tbl:WFUQ} shows a comparison of the marginal uncertainty quantification when the forecasts are uncalibrated (using the ensemble standard deviation to derive the uncertainty estimates \citep{mcd17}) and calibrated using the aforementioned approach. From this table it can be seen that, when using the uncalibrated estimates, the marginal uncertainty is incorrectly quantified which leads to a worse coverage of the nominal 95\%, 90\%, and 80\% prediction intervals (PIs). Here the average discrepancy in coverage from the uncalibrated estimates across all of the 100 subsampled location is 8.3\% across the three intervals. When using the uncertainty estimates produced by the calibration approach it is clearly seen from the table how the marginal uncertainty is now more correctly quantified and does capture a relatively appropriate amount of the true data within the specified intervals. More specifically, the calibrated 95\% PI now capture a median of 95.0\% of the true data across both space and time for all locations. When using the calibrated intervals, there is only a coverage discrepancy of 3.3\% across the three intervals, thus further exemplifying the improvement over the uncalibrated approach.

\begin{table}[tb!]
\begin{centering}
\begin{tabular}{||c|c|c||} 
\hline
Prediction Interval & Uncalibrated & Calibrated \\[1ex] 
\hline\hline
95\% & 100.0 (0.0) & 95.0 (1.0) \\ 
\hline
90\% & 100.0 (1.0) & 90.0 (2.0) \\ 
\hline
80\% & 90.0 (3.0) & 90.0 (2.0) \\
\hline
\end{tabular}
\caption{Uncertainty quantification methods compared in terms of their nominal 95\% PI coverage on water field data. The ensemble-based (uncalibrated) approach is outlined in \cite{mcd17} and the quantile regression (calibrated) approach is the method detailed in \cite{bonas23b}. The median across the 100 subsampled locations is reported with the IQR in parenthesis.}
\label{tbl:WFUQ}
\end{centering}
\end{table}

\section{Conclusion}\label{sec:conclusion}

\quad In this work we have introduced a new approach to forecasting boundary layer velocity by combining two reservoir computing approaches while also incorporating a physical constraints with a PDE penalization. The proposed model is able to achieve superior forecasting skills while (partly) complying with physical constraints, as well as accurate uncertainty quantification. While the proposed approach has been applied to a turbulent flow in a controlled experimental setting, the principles underpinning the proposed methodology are applicable to any spatio-temporal data set for which additional context in the form of a PDE (or systems of PDEs) is available. 

Our proposed approach comprises of two main innovations. Firstly, we combine two deep reservoir computing methods, Echo State Networks and Liquid State Machines, one relying on continuous signals, the other one on spike trains. Secondly, inference is not performed with a standard bias-variance tradeoff approach such as ridge regression, but with a physics-informed penalty stemming from a PDE, which is assumed to be known and at least partially descriptive of the underlying process dynamics. The resulting physics-informed deep double reservoir model significantly outperforms the other reference models presented, such as ARIMA and LSTMs, in terms of their forecasting skills in both the case of the simulated Burgers' equations \citep{burger48} data in Section \ref{sec:simstudy} as well as the application detailed in Section \ref{sec:Data} with corresponding results in Section \ref{sec:Appl}. Crucially, the forecast is also more compliant with some key physical properties such as energy dissipation (for the simulation study) as well as mass conservation and realism of the fluctuating velocity signal (for the application). In order to calibrate the uncertainty propagated by the two reservoir computing approaches, we further augment the ensemble predictions with the calibration approach and showed how this results in accurate uncertainty quantification.

The proposed method is based on on combining two reservoir computing approaches, whose flexibility stems from the ability to capture nonlinear dynamics without requiring very large training data as standard recurrent NNs. More importantly from a statistics perspective, this work contributes to a model-based view of this class of methods, which have been originally envisioned as a possible solution to the vanishing/exploding gradient when performing inference on a recurrent NNs. While reservoir computing models do indeed sidestep the issue of the numerical instabilities in gradient calculation via cross-validation on a small parameter space, from a statistician perspective this is only the consequence on the use of prior in the weight matrices. As such, while the introduction of additional uncertainty is necessary, our stance is that such uncertainty can and should be quantified in the same fashion as any other statistical model.

The addition of a physical constrain in the model via optimization of a modified, PDE-driven penalty represent a powerful, simple and appealing unifying framework for problems ranging from data-rich/context-poor to data-poor/context-rich. It does, however, come with associated issues one should be aware of. Since the penalization is adding a bias in the inference towards what is believed to be a more physically consistent process, the physics-based forecast is bound to have inferior short-term point-predictive skills (measured as MSE). In this work, we have shown that this loss of predictability is counter-balanced with some degree of preservation of additional physical properties in the forecast, such as loss of energy and mass conservation, as well as improved long lead forecasting in the simulation study. Clearly, which properties are to preserved is a very application dependent question and contextualization of the problem is necessary.

While the proposed approach has shown improved forecasting skills in a turbulent boundary layer in a controlled laboratory, this work represents only the first step towards an operational use on a real case scenarios such as in the atmosphere. While the experimental data used here represents a scale model of the types of turbulent motions found in the atmosphere (and whose prediction could assist in engineering fields such as wind energy production), in reality, many if not all the simplifying assumptions of the Navier-Stokes equation which allowed us to reduce the problem to a scalar PDE will have to be relaxed. As such, a real-case penalty will need to be a system of PDEs more akin to the original equations, possibly coupled with those associated to the pressure field, with a considerable increase in the computational overhead. What will be the simplifying assumptions on the PDE penalty that will allow the inference to still be computationally feasible is expected to be a challenging future area of research.


\begin{center}
\textbf{\large Supplementary Material}
\end{center}

\noindent We provide the \texttt{R} code to apply the DESN, DDRN, PIDESN, and PIDDRN forecasting approaches mentioned in Sections \ref{sec:methods} on a simulated Burgers' equation \cite{burger48} data with 30 spatial locations and 500 time steps. This code and data can be found in the following GitHub repository: \url{github.com/Env-an-Stat-group/25.Bonas.JASA}.

\bibliographystyle{chicago}
\bibliography{references}

\end{document}